\documentclass{mrow-ijmpe}
\usepackage{graphicx}

\newcommand{\be}{\begin{equation}}
\newcommand{\ee}{\end{equation}}
\newcommand{\ba}{\begin{eqnarray}}
\newcommand{\ea}{\end{eqnarray}}
\newcommand{\ban}{\begin{eqnarray*}}
\newcommand{\ean}{\end{eqnarray*}}

\begin{document}

\markboth{Maj \& Mr\'owczy\'nski}{Towards reliable calculations of the correlation function}

\catchline{}{}{}{}{}

\title{TOWARDS RELIABLE CALCULATIONS \\ OF THE CORRELATION 
FUNCTION\footnote{lecture given at XIth International Workshop
on {\it Correlation and Fluctuation in Multiparticle Production}, 
Hangzhou, China, November 21-24, 2006}}

\author{Rados\l aw Maj}

\address{Institute of Physics, \'Swi\c etokrzyska Academy \\
ul. \'Swi\c etokrzyska 15, PL - 25-406 Kielce, Poland \\
radmaj@pu.kielce.pl}

\author{Stanis\l aw Mr\' owczy\' nski\footnote{speaker}}
\address{Institute of Physics, \'Swi\c etokrzyska Academy \\
ul. \'Swi\c etokrzyska 15, PL - 25-406 Kielce, Poland \\
and So\l tan Institute for Nuclear Studies \\
ul. Ho\.za 69, PL - 00-681 Warsaw, Poland \\
mrow@fuw.edu.pl}

\maketitle

\begin{history}
\end{history}

\begin{abstract}

The correlation function of two identical pions interacting via 
Coulomb potential is computed for a general case of anisotropic 
particle's source of finite life time. The effect of halo is taken 
into account as an additional particle's source of large spatial 
extension. Due to the Coulomb interaction, the effect of halo is 
not limited to very small relative momenta but it influences the 
correlation function in a relatively large domain. The relativistic 
effects are discussed in detail and it is argued that the calculations 
have to be performed in the center-of-mass frame of particle's pair 
where the (nonrelativistic) wave function of particle's relative motion 
is meaningful. The Bowler-Sinyukov procedure to remove the Coulomb 
interaction is tested and it is shown to significantly underestimate 
the source's life time. 

\end{abstract}


\section{Introduction}


The correlation functions of two particles with `small' relative
momenta provide information about space-time characteristics of
particle's sources in high-energy nucleus-nucleus collisions, see the
review articles\cite{Wiedemann:1999qn,Heinz:1999rw,Lisa:2005dd}.
Within the standard `femtoscopy' method, one obtains parameters of 
a particle's source, which is usually called the fireball, comparing 
the experimental correlation functions to the theoretical ones which 
are calculated in a given model. Such an analysis can be performed 
for pairs of non-identical or identical particles. In the former case,
the correlation appears due to inter-particle interaction while in
the latter one the interaction is combined with the effects of
quantum statistics. Since we usually deal with electrically charged
particles, observed two-particle correlations are influenced, if not
dominated, by the Coulomb interaction. The effect of the Coulomb
force is treated as a correction and it is usually eliminated from
the experimental data by means of the so-called Bowler-Sinyukov
procedure\cite{Bowler:1991vx,Sinyukov:1998fc}. And then, the
correlation function, which is obtained in such a way from
experimental data, is compared with the theoretical correlation
function of two non-interacting particles. The latter one is
computed for a given source function - a distribution of the
emission points - parameterized usually in a gaussian form. The
comparison with the experimental data provides parameters of the
source function.

Within the method of imaging\cite{Brown:1997ku,Brown:2005ze}, one
obtains the source function not referring to its specific
parametrization but directly inverting the functional dependence of
the correlation function on the source function. The procedure of
inversion involves the effect of quantum statistics as well as that
of inter-particle interaction. The method provides essentially model
independent information on the source space-time sizes but modeling
is still needed to deduce the source life time which is coupled to 
the spatial size parameters. The method of one-dimensional imaging
was successfully applied to experimental 
data\cite{Panitkin:2001qb,Chung:2002vk} but the three-dimensional
imaging is technically very complex and cumbersome\cite{Brown:2005ze}.
However, very first results of the three-dimensional imaging
applied to NA49 data are presented in these proceedings\cite{Chung2006}.

The femtoscopy was applied to a large volume of experimental data on
nucleus-nucleus collisions at SPS energy\cite{Heinz:1999rw}. The
spatial size of particle's sources appeared to be comparable to the
expected size a fireball created in nucleus-nucleus collisions while
the emission time of particles was significantly shorter. It was
predicted that at RHIC energies the emission time would be
significantly longer due to the long lasting hydrodynamic evolution
of the system created at the early stage of nucleus-nucleus
collisions\cite{Rischke:1995cm,Rischke:1996em}. To a big surprise
the experimental data obtained at 
RHIC\cite{Adler:2001zd,Adcox:2002uc,Adler:2004rq,Adams:2004yc} show 
a very little, if any, change of the space-time characteristics of 
a fireball when compared to the SPS data. In particular, the emission 
time of particles appeared to be as short as 1 fm/c. This surprising 
result, which is now known as the `HBT  Puzzle'\cite{Gyulassy:2001zv,Pratt:2003ij},
demonstrates that either we do not really understand the dynamics of
relativistic heavy-ion collisions or the principles of the
femtoscopy are doubtful. In any case, the femtoscopy method need to
be examined.

In this presentation there is given a preliminary account of our study 
which is aimed to check three aspects of the standard method. Relativistic 
effects are, in our opinion, not satisfactory incorporated, as the 
non-relativistic wave function of two interacting particles is often 
implicitly treated as a Lorentz scalar. It should be stressed that 
a transformation law of a wave function under Lorentz transformations 
is unknown\footnote{Using the Bethe-Salpeter equation, it has been 
recently shown that the hydrogen atom wave function experiences the 
Lorentz contraction\cite{Jarvinen:2004pi} under the Lorentz boost.}. 
However, the observed correlation functions are significantly different 
from unity for a small relative momenta when the relative motion of
particles is non-relativistic. Therefore, it is fully legitimate to 
use the non-relativistic wave function in the center-of-mass frame 
of two particles. However, it requires an explicit transformation of the 
source function to the center-of-mass frame. In Sec.~\ref{sec-relat} 
the relativistic effects are discussed in detail and it is shown that 
a proper treatment of them leads to a numerically different result 
when compared with the standard procedure.

The correlation function of two identical non-interacting bosons is
expected to be equal to 2 for vanishing relative momentum of the two
particles. The correlation function extracted from experimental data
by means of the procedure, which is supposed to remove the Coulomb
interaction, does not posses this property. The correlation function
at zero relative momentum rather equals $1 + \lambda$ with $ 0 <
\lambda < 1$. This fact can be explained referring to the concept
of halo\cite{Nickerson:1997js}. It assumes that only a fraction of 
observed particles, which equals $\lambda$, comes from the fireball 
while the rest originates from the long living resonances. Then, we 
have two sources of particles: the fireball, which is rather small, 
and the halo with the radius given by the distance traveled by long 
living resonances. The complete correlation function, which includes
particles from the fireball and the halo, equals 2 at exactly
vanishing relative momentum. However, the quantum statistical
correlation of two particles coming from the halo occurs at the
relative momentum which is as small as the inverse radius of the
halo. Since experimental momentum resolution is usually much
poorer and such small relative momenta are not accessible, the
correlation function is claimed to be less than 2 for {\em effectively}
vanishing relative momentum. While the effect of halo is commonly
believed to resolve the problem, the concept has been never examined
for the particles which experience Coulomb interaction. In
Sec.~\ref{sec-halo} we show that the effect of halo is not limited
to very small relative momenta but due to the Coulomb repulsion the
correlation function is modified for larger momenta.

The Bowler-Sinyukov correction procedure, which is used to eliminate
the Coulomb interaction from the experimental data, assumes that the
Coulomb effects can be factorized out. The correction's factor is 
calculated for a particle's source which is spherically symmetric 
and has zero life time. We find such an approach rather inconsistent 
as the objective is to determine the source spatial parameters not 
assuming any source symmetry. The Bowler-Sinyukov procedure is 
examined in Sec.~\ref{sec-B-S} where we also comment on a similar 
test of the procedure performed in the paper\cite{Kisiel:2006is}.

We use the natural units, where $c = \hbar = 1$, and the metric
convention $(+,-,-,-)$.


\section{Definition}
\label{sec-def}


The correlation function $C(\mathbf{p}_{1}, \mathbf{p}_{2})$
of two particles with momenta $\mathbf{p}_{1}$ and $\mathbf{p}_{2}$
is defined as
$$
C({\bf p}_1, {\bf p}_2)
=\frac{\frac{dN}{d{\bf p}_1 d{\bf p}_2}}
{\frac{dN}{d{\bf p}_1}\frac{dN}{d{\bf p}_2}} \;,
$$
where $\frac{dN}{d\mathbf{p}_1 d\mathbf{p}_2}$ and
$\frac{dN}{d\mathbf{p}_1}$ is, respectively, the two- and
one-particle momentum distribution. The correlation function can be
written down in a Lorentz covariant form
\be
\label{def-cov}
C({\bf p}_1, {\bf p}_2)=
\frac{E_1 E_2\frac{dN}{d\mathbf{p}_1 d\mathbf{p}_2}}
{E_1\frac{dN}{d\mathbf{p}_1} E_2\frac{dN}{d\mathbf{p}_2}} \;,
\ee
where $E\frac{dN}{d^3\mathbf{p}}$ is the Lorentz invariant
distribution.

The covariant form (\ref{def-cov}) shows that the correlation function 
is  Lorentz scalar which can be easily transformed from one to 
another reference frame. If the particle four-momenta, which 
are on mass-shall, transform as ${\bf p}_i \rightarrow {\bf p'}_i$ 
with $i=1,2$, the transformed correlation function equals
$$
C'(\mathbf{p'}_{1}, \mathbf{p'}_{2})=
C(\mathbf{p}_1(\mathbf{p'}_{1}), \mathbf{p}_2(\mathbf{p'}_{2})).
$$


\section{Nonrelativistic Koonin formula}
\label{sec-koonin}


Within the Koonin model\cite{Koonin:1977fh}, the correlation
function $C$ can be expressed in the source rest frame as
\be 
\label{Koonin} 
C(\mathbf{p}_1,\mathbf{p}_2) = 
\int d^3r_1 dt_1 d^3r_2 dt_2 D(\mathbf{r}_1,t_1) \:
D(\mathbf{r}_2,t_2) \: |\Psi(\mathbf{r}'_1,\mathbf{r}'_2)|^2 \;, 
\ee
where $\mathbf{r}'_i \equiv \mathbf{r}_i+\mathbf{v}_i t_i$,
$\Psi(\mathbf{r}'_1,\mathbf{r}'_2)$ is the wave function of 
the two particles and $D(\mathbf{r},t)$ is the single-particle 
source function which gives the probability to emit the particle 
from the space-time point $(t,\mathbf{r})$. The source function
is normalized as 
\be
\label{norma}
\int d^3r \, dt \, D(\mathbf{r},t)=1 \;.
\ee

After changing the variables ${\bf r}' \leftrightarrow {\bf r}$, the 
correlation function can be written in the form
$$
C(\mathbf{p}_{1}, \mathbf{p}_{2}) = 
\int d^3 r_1 dt_1 d^3 r_2 dt_2 
D(\mathbf{r}_1-\mathbf{v}_1t_1,t_1) \:
D(\mathbf{r}_2-\mathbf{v}_2t_2,t_2)
|\Psi(\mathbf{r}_1,\mathbf{r}_2)|^2 \;.
$$

Now, we introduce the center-of-mass coordinates
\ban
\begin{array}{cc}
\mathbf{r}=\mathbf{r}_2-\mathbf{r}_1, &
\mathbf{R}=\frac{1}{M}(m_1\mathbf{r}_1+m_2\mathbf{r}_2), 
\\[2mm]
t=t_2-t_1, & T=\frac{1}{M}(m_1 t_1+m_2 t_2),
\\[2mm]
\mathbf{q}= \frac{1}{M}(m_2 \mathbf{p}_1 - m_1\mathbf{p}_2),&
\mathbf{P}=\mathbf{p}_1+\mathbf{p}_2,
\end{array}
\ean
where $M \equiv m_1+m_2$. Using the center-of-mass
variables, one gets
\be 
\label{Koonin-r} 
C(\mathbf{q})= 
\int d^3r\:dt\: D_r(\mathbf{r-v}t,t)
|\varphi_\mathbf{q}(\mathbf{r})|^2, 
\ee
where the `relative' source function is defined as
\be
\label{source-relat}
D_r(\mathbf{r},t) \equiv \int d^3R \, dT \:
D(\mathbf{R}-\frac{m_2}{M}\mathbf{r},T-\frac{m_2}{M}t) \:
D(\mathbf{R}+\frac{m_1}{M}\mathbf{r},T+\frac{m_1}{M}t) .
\ee
To get Eq.~(\ref{Koonin-r}), the wave function was factorized as
$$
\Psi(\mathbf{r}_1,\mathbf{r}_2)=
e^{i\mathbf{P}\mathbf{R}}\varphi_\mathbf{q}(\mathbf{r})
$$
with $\varphi_\mathbf{q}(\mathbf{r})$ being the wave function of 
the relative motion in the center-of-mass frame. Deriving
Eq.~(\ref{Koonin-r}), it has been assumed that the particles move with
equal velocities i.e. $\mathbf{v}_1=\mathbf{v_2}=\mathbf{v}$ which
requires, strictly speaking, ${\bf q} = 0$. However, one observes
that $|{\bf v}_1 - {\bf v}_2 | \ll |{\bf v}_i|$ if 
$|{\bf q} | \ll \mu |{\bf p}_i|/m_i$ where $\mu \equiv m_1 m_2/M$. 
Thus, the approximation $\mathbf{v}_1=\mathbf{v_2}$ holds for a 
sufficiently small center-of-mass momentum.

We choose the gaussian form of the single-particle source function
$D({\bf r},t)$ 
\be
\label{gauss}
D(\mathbf{r},t)=\frac{1}{4\pi^2 R_x R_y R_z \tau} \:
\exp\Big[-\frac{x^2}{2R_x^2} - \frac{y^2}{2R_y^2}
- \frac{z^2}{2R_z^2} - \frac{t^2}{2\tau^2}\Big],
\ee
where ${\bf r}=(x,y,z)$ and the parameters $\tau$, $R_x$, $R_y$
and $R_z$ characterize the life time and sizes of the source.
Specifically, the parameters $\tau$ and $R_x$ give, respectively,
$$
\tau^2 = \langle t^2 \rangle 
\equiv \int d^3r \: dt\: t^2 D(\mathbf{r},t) \;,
\;\;\;\;\;\;\;\;\;
R_x^2=\langle x^2\rangle 
\equiv \int d^3r \: dt\: x^2 D(\mathbf{r},t) \;.
$$

The relative source function computed from Eq.~(\ref{source-relat})
with the single-particle source (\ref{gauss}) is
\be
\label{gauss-relat}
D_r(\mathbf{r},t)=\frac{1}{16\pi^2 R_x R_y R_z \tau} \:
\exp\Big[-\frac{x^2}{4R_x^2}-\frac{y^2}{4R_y^2}-\frac{z^2}{4R_z^2}-
\frac{t ^2}{4\tau^2} \Big].
\ee
We note that the particle's masses, which are present in the definition
(\ref{source-relat}), disappear completely in the formula (\ref{gauss-relat}).
This is the feature of the gaussian parameterization (\ref{gauss}). 

In the case of non-interacting identical bosons, the two-particle
symmetrized wave function is
$$
\Psi(\mathbf{r}_1,\mathbf{r}_2)= \frac{1}{\sqrt{2}}[e^{i\mathbf{p}_1\mathbf{r}_1+\mathbf{p}_2\mathbf{r}_2}
+ e^{i\mathbf{p}_2\mathbf{r}_1+\mathbf{p}_1\mathbf{r}_2}]=
\frac{1}{\sqrt{2}}[e^{i\mathbf{q}\mathbf{r}}+
 e^{-i\mathbf{q}\mathbf{r}}]e^{i\mathbf{PR}}.
$$
It gives the modulus square of the wave function of relative motion 
$|\varphi_{\bf q}({\bf r})|^2=1+\cos{(2{\bf q}{\bf r})}$ which in turn
provides the correlation function equal to
\be 
\label{corr-free}
C({\bf q}) = 1 + \exp \big[ - 4 \big(
\tau^2({\bf qv})^2 + R_x^2 q_x^2 + R_y^2 q_y^2 + R_z^2 q_z^2 \big)
\big] \;,
\ee
where ${\bf q} \equiv (q_x,q_y,q_z)$. We note that the `cross terms' such 
as $q_x q_z$ do not show up as the source function (\ref{gauss}) obeys 
the mirror symmetry $D({\bf r},t) = D(-{\bf r},t)$. We also note that 
${\bf q}$ often denotes the relative momentum ${\bf p}_1 - {\bf p}_2$ not 
the center-of-mass momentum, which for equal mass nonrelativistic particles
equals $\frac{1}{2}({\bf p}_1 - {\bf p}_2)$, and then, the factor 4 does 
not show up in the correlation function (\ref{corr-free}) of identical 
free bosons. However, we believe that using the center-of-mass momentum 
is physically better motivated.


\section{Relativistic formulations}
\label{sec-relat}


There are two natural ways to `relativize' the Koonin formula (\ref{Koonin}).
The first one provides explicitly Lorentz covariant correlation function but
it is applicable only for the non-interacting particles. The second one holds
only in a specific reference frame but it is applicable for interacting 
particles as well. Below, we consider the two methods.  We start, however, 
with the discussion of the Lorentz covariant form of the source function.

\subsection{Lorentz covariant source function}

The Lorentz covariant form of the gaussian parameterization
of the source function (\ref{gauss}) is
\be
\label{source-cov}
D(x)=\frac{\sqrt{{\rm det}\Lambda}}{4\pi^2} \;
{\rm exp} [-\frac{1}{2}x_\mu \Lambda^{\mu\nu}x_\nu],
\ee
where $x^\mu$ is the position four-vector and $\Lambda^{\mu\nu}$
is the Lorentz tensor characterizing the source which in the
source rest frame is
\be
\label{source-matrix}
\Lambda^{\mu\nu}=\left[\begin{array}{cccc}
\frac{1}{\tau^2} & 0 & 0 & 0  \\
0 & \frac{1}{R_{x}^2} & 0 & 0  \\
0 & 0 & \frac{1}{R_{y}^2} & 0  \\
0 & 0 & 0 & \frac{1}{R_{z}^2}  \\
\end{array}\right].
\ee
The source function as written in Eq.~(\ref{source-cov}) obeys
the normalization condition (\ref{norma}) not only for the diagonal
matrix $\Lambda$ but for non-diagonal as well. 

The source function (\ref{source-cov}) is evidently the Lorentz 
scalar that is
$$
D'(x') = \frac{\sqrt{{\rm det} \Lambda'}}{4 \pi^2}
\exp{[-\frac{1}{2} x'_\mu \Lambda'^{\mu\nu}x'_\nu]}
= \frac{\sqrt{{\rm det} \Lambda}}{4 \pi^2}
\exp{[-\frac{1}{2} x_\mu \Lambda^{\mu\nu}x_\nu]}
= D(x) \;,
$$
where $x'_\mu = L_{\mu}^{\;\;\nu}x_\nu$
and $\Lambda'^{\mu\nu} = L_{\;\;\sigma}^{\mu}\Lambda^{\sigma\rho}L_{\rho}^{\;\;\nu}$
with $L_{\sigma}^{\mu}$ being the matrix of Lorentz transformation. 
We note that ${\rm det} \Lambda' = {\rm det}L \: {\rm det} \Lambda
\: {\rm det}L^{-1} = {\rm det} \Lambda$. 

The covariant relative source function (\ref{gauss-relat}) 
is given by
\be 
\label{source-rel-cov} 
D_r(x) = \frac{\sqrt{{\rm det} \Lambda}}{16\pi^2} 
\exp{[-\frac{1}{4} x_\mu \Lambda^{\mu\nu}x_\nu]}\;. 
\ee

\subsection{Explicitly covariant `relativization'}
\label{sub-sec-exp-cov}

As follows from Eq.~(\ref{def-cov}), the correlation function is
a Lorentz scalar. Therefore, the Koonin formula (\ref{Koonin}) can 
be `relativized' demanding its Lorentz covariance. Let us write the 
formula
\be
\label{explicit}
C(p_1,p_2) = \int d^4x_1d^4x_2 D(x_1) \:
D(x_2)|\Psi(x_1,x_2)|^2,
\ee
where $p_i$ and $x_i$ is, respectively, the four-momentum and
four-position. Since the source function $D(x)$ and the four-volume
element $d^4x_i$ are both the Lorentz scalars, the whole formula 
(\ref{explicit}) is covariant if the wave function $\Psi(x_1,x_2)$ 
is covariant as well. In the case of non-interacting bosons the 
relativistic wave function $\Psi(x_1,x_2)$
is
\be
\label{rel-wave-fun}
\Psi(x_1,x_2)= \frac{1}{\sqrt{2}}
(e^{ip_1 x_1 + i p_2 x_2} + e^{ip_1 x_2 + i p_2 x_1}).
\ee
As the function depends on the scalar products of two four-vectors,
this is the Lorentz scalar. We note that the function 
(\ref{rel-wave-fun}) depends on two time arguments. 

Our further considerations are limited to pairs of identical 
particles and thus, we introduce the relative coordinates as
\ba
\label{relative}
\begin{array}{cc}
x = x_2-x_1, & X = \frac{1}{2}(x_1 + x_2), 
\\[2mm]
q = \frac{1}{2}(p_1 - p_2),& P = p_1 + p_2.
\end{array}
\ea
We note that in the non-relativistic treatment the three-vectors
${\bf r}$ and ${\bf q}$, which are given by the four-vectors 
$x=(t,{\bf r})$ and $q=(q_0, {\bf q})$, correspond to the  
inter-particle separation and the particle's momentum in the 
center-of-mass of the particle pair. This is, however, not
the case in the relativistic domain. To get the center-of-mass 
variables, the four-vectors need to be Lorentz transformed. 

With the variables (\ref{relative}), the wave function 
(\ref{rel-wave-fun}) equals
$$
\Psi(x,X)=\frac{1}{\sqrt{2}}(e^{iqx}+e^{-iqx})e^{-iPX} \;,
$$
and the correlation function is found in the form
$$
C(q)=1+\exp[-4q_\mu(\Lambda^{\mu\nu})^{-1}q_\nu] \;,
$$
which is explicitly Lorentz covariant. For the source matrix 
(\ref{source-matrix}), the correlation function equals
\be
\label{corr-free-relat}
C(q)= 1+ \exp \big[-4
(q_0^2\tau^2 + q_x^2R_x^2 + q_y^2R_y^2 + q_z^2R_z^2)\big] \;.
\ee
If $|{\bf q}| \ll |{\bf p}_i|$ with $i=1,2$, then 
$q_0 \approx {\bf qv}$, and the correlation function 
(\ref{corr-free-relat}) exactly coincides with the non-relativistic 
expression (\ref{corr-free}). This coincidence is not completely
obvious as the time variables enter differently in the Koonin
formula (\ref{Koonin}) and in the covariant one (\ref{explicit}).

Let us consider the correlation function in the center-of-mass
frame of the particle's pair. We assume that the velocity of 
the center-of-mass frame in the source rest frame is along the 
axis $x$. Then,  ${\bf v}=(v,0,0)$ and $q_0 \approx {\bf qv}=q_xv$. 
The correlation function (\ref{corr-free-relat}), which holds in 
the source rest frame, equals
\be 
\label{corr-free-relat-vx}
C(\mathbf{q})=1+ \exp
\big[-4\big( (v^2\tau^2 + R_x^2)q_x^2 
+ R_y^2 q_y^2+ R_z^2 q_z^2 \big)\big]\;.
\ee
As seen, the effective source radius in the direction $x$ is
$\sqrt{R_x^2 + v^2\tau^2}$. We now transform the source function 
to the center-of-mass frame where the quantities are labeled with 
the index $*$. The center-of-mass source matrix (\ref{source-matrix}), 
which is computed as 
$$
\Lambda^{\mu\nu}_* = L_{\;\;\sigma}^{\mu}
\Lambda^{\sigma\rho} L_{\rho}^{\;\;\nu},
$$
where
$$
L_{\;\;\sigma}^{\mu}=\left[\begin{array}{cccc}
\gamma & -v\gamma & 0 & 0  \\
-v\gamma & \gamma & 0 & 0  \\
0 & 0 & 1 & 0  \\
0 & 0 & 0 & 1  \\
\end{array}\right] 
$$
with $\gamma \equiv (1-v^2)^{-1/2}$, equals
\ba
\label{source-matrix-CM}
\Lambda^{\mu\nu}_*=\left[\begin{array}{cccc}
\gamma^2(\frac{1}{\tau^2}+\frac{v^2}{R_{x}^2}) & 
- \gamma^2v(\frac{1}{\tau^2}+\frac{1}{R_{x}^2}) & 0 & 0  \\
-\gamma^2v(\frac{1}{\tau^2}+\frac{1}{R_{x}^2}) & 
\gamma^2(\frac{v^2}{\tau^2}+\frac{1}{R_{x}^2}) & 0 & 0  \\
0 & 0 & \frac{1}{R_{y}^2} & 0  \\
0 & 0 & 0 & \frac{1}{R_{z}^2}  \\
\end{array}\right] \;.
\ea
Then, the correlation function in the center-of-mass frame is found as
\ba 
\label{corr-free-relat-vx-CM} 
C(q_*) &=& 1 + 
\exp [-4q_{*\mu}(\Lambda^{\mu\nu}_*)^{-1}q_{*\nu} ]
\\ [2mm] \nonumber 
&=& 
1 + \exp \big[-4 \big( \gamma^2(v^2\tau^2 + R_x^2)q_{*x}^2 
+ R_y^2 q_{*y}^2+ R_z^2 q_{*z}^2\big)\big]\;.
\ea 
As seen, the effective source radius along the direction of
the velocity is elongated, not contracted as one can naively
expect, by the factor $\gamma$.

\subsection{Non-covariant relativization}

The quantum mechanical description of two relativistic interacting
particles faces serious difficulties. The problem is greatly
simplified when the relative motion of two particles is
non-relativistic (with the center-of-mass motion being fully
relativistic). Since the correlation functions usually differ from
unity only for small relative momenta of particles, it is reasonable
to assume that the relative motion is non-relativistic. We further 
discuss the correlation functions taking into account the
relativistic effects of motion of particles with respect to the
source but the particle's relative motion is treated
non-relativistically. In such a case, the wave function of relative
motion is a solution of the non-relativistic Schr\"odinger equation.
Thus, we compute the correlation function directly from the Koonin
formula (\ref{Koonin}) but we use it in the center-of-mass frame
of the pair. For this reason we first transform the source function
to this frame and then, after performing the integrations over 
$x_1$ and $x_2$, we transform the whole correlation function,
which is known to be a Lorentz scalar, to the source rest frame.

As already stressed, we compute the correlation function in the 
center-of-mass frame of the pair and we use the relative variables 
(\ref{relative}). Since the source function has the gaussian form 
(\ref{source-cov}) (with the non-diagonal matrix $\Lambda$), the 
integration over $X$ can be easily performed and the correlation 
function equals
\be
\label{Koonin-relat}
C({\bf q}_*) = \int
d^4x_* \: D_r(x_*) \:
|\varphi_{{\bf q}_*}({\bf r'}_*)|^2,
\ee
where $D_r(x_*)$ is  the relative source function 
(\ref{source-rel-cov}) and $\varphi_{{\bf q}_*}({{\bf r}'}_*)$ 
with ${{\bf r}'}_*\equiv {\bf r}_* + {\bf v}_*t_*$ is the 
non-relativistic wave function of relative motion.

\begin{figure}[t]
\begin{minipage}{6.0cm}
\includegraphics*[width=5.8cm]{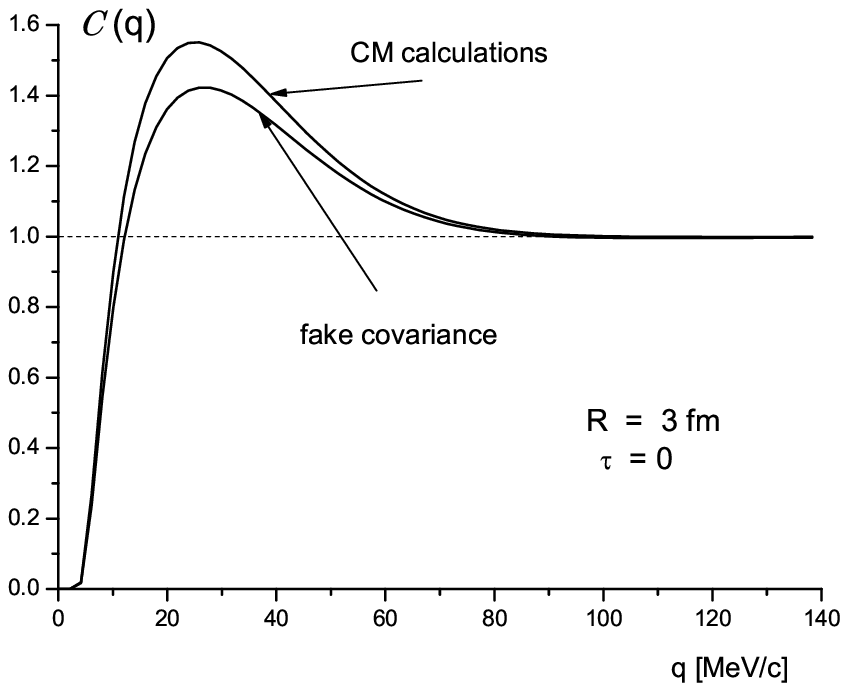}
\vspace{-5mm}
\caption{The Coulomb correlation function computed directly 
in the source rest frame (`fake covariance') and the correlation 
function computed in the center-of-mass frame of the pair and 
then transformed to the source rest frame (`CM calculations').}
\label{fig-cm-fake}
\end{minipage}\hspace{5mm}
\begin{minipage}{6.0cm}
\vspace{-7mm}
\includegraphics*[width=5.8cm]{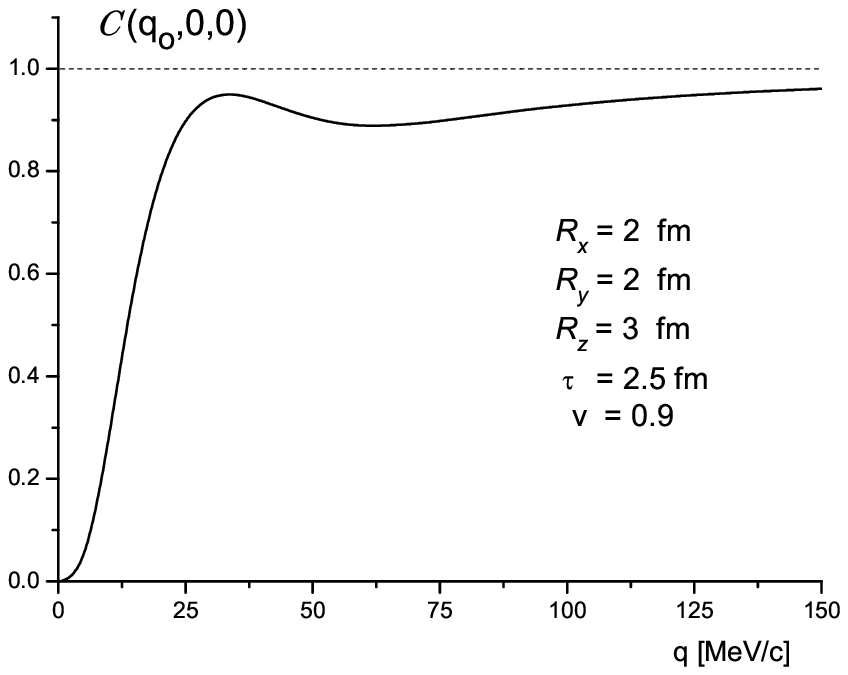}
\vspace{-4mm}
\caption{The Coulomb correlation function $C(q_o,0,0)$ 
as a function of $q_o$ for the source parameters:
$R_x = 2$~fm, $R_y = 2$~fm, $R_z = 3$~fm, $\tau=2.5$~fm and
${\bf v} = (0.9,0,0)$.}
\label{fig-out}
\end{minipage}
\end{figure}

Although our aim is to compute the correlation functions of
interacting particles, we start the discussion with the free
identical bosons for the sake of comparison with the results of
the previous section \ref{sub-sec-exp-cov} where the covariant
`relativisation' was presented. We again assume that 
${\bf v} = (v,0,0)$, and then the source matrix is given by 
Eq.~(\ref{source-matrix-CM}). The correlation function, which 
follows from Eq.~(\ref{Koonin-relat}), exactly coincides with 
the formula (\ref{corr-free-relat-vx-CM}). To get the correlation 
function in the source rest frame, one performs the Lorentz 
transformation and obtains the formula (\ref{corr-free-relat-vx}). 
Thus, the two ways of `relativization' give the same result for 
non-interacting particles. This is not quite trivial as the time 
dependence of the Koonin formula (\ref{Koonin-relat}) and of 
the explicitly covariant one (\ref{explicit}) is rather different.

In the following sections we compute the correlation function of 
identical pions interacting due to Coulomb force, using the Koonin formula 
(\ref{Koonin-relat}). Thus, we first compute the correlation function 
in the center-of-mass frame of the pair, and then, we transform it to 
the source rest frame.  The correlation function of interacting particles 
is often calculated following the explicitly covariant method described in 
Sec.~\ref{sub-sec-exp-cov}. And then, the non-relativistic wave function 
is treated as a Lorentz scalar function. Sometimes one argues that it 
must be a Lorentz scalar to guarantee that the right-hand side of 
Eq.~(\ref{explicit}) is the Lorentz scalar. We believe that such a 
procedure is incorrect and we refer to it as to `fake covariance'. 
In principle, one can argue that the function $\Psi(x_1,x_2)$ from 
Eq.~(\ref{explicit}) has to be a Lorentz scalar but it is unjustified 
to identify it with the non-relativistic wave function. The wave function 
is well defined in the center-of-mass of the pair and in the reference 
frames which move non-relativistically with respect to it. But properties 
of the wave function under Lorentz transformations are unknown, and thus, 
there is no reliable way to transform it from one frame to another. We 
can also put it differently. The correlation function as defined by 
Eq.~(\ref{def-cov}) is certainly a Lorentz scalar but there is no 
guarantee that the theoretical model (\ref{Koonin}) gives the correlation 
function which is a Lorentz scalar. However, we expect that the model 
works well in the center-of-mass frame of the pair where the 
non-relativistic wave function is well defined. Thus, we can compute 
the correlation function in this frame and then, we can transform it 
to an arbitrary frame, knowing that the correlation function is a Lorentz 
scalar. In this way, the non-covariant procedure circumvents the problem 
of unknown transformation properties of the non-relativistic wave function.

Although the problem of transformation properties looks somewhat academic, 
it leads to a numerically significant effect. In Fig.~\ref{fig-cm-fake}
we show the correlation function of two identical charged pions
computed in the center-of-mass frame of the pair and then transformed 
to the source rest frame. The function is compared to the correlation
function which is directly computed in the source rest frame, treating
the wave function as a Lorentz scalar. The source is assumed here 
to be spherically symmetric with $R = 2$ fm and $\tau =0$. The pair 
velocity with respect to the source chosen to be $v=0.9$. (Details of
the calculations of the Coulomb correlation functions are discussed in 
the next section.) As seen, the assumption of `fake covariance' 
noticeably distorts the correlation function.

\begin{figure}[t]
\begin{minipage}{6.0cm}
\includegraphics*[width=5.8cm]{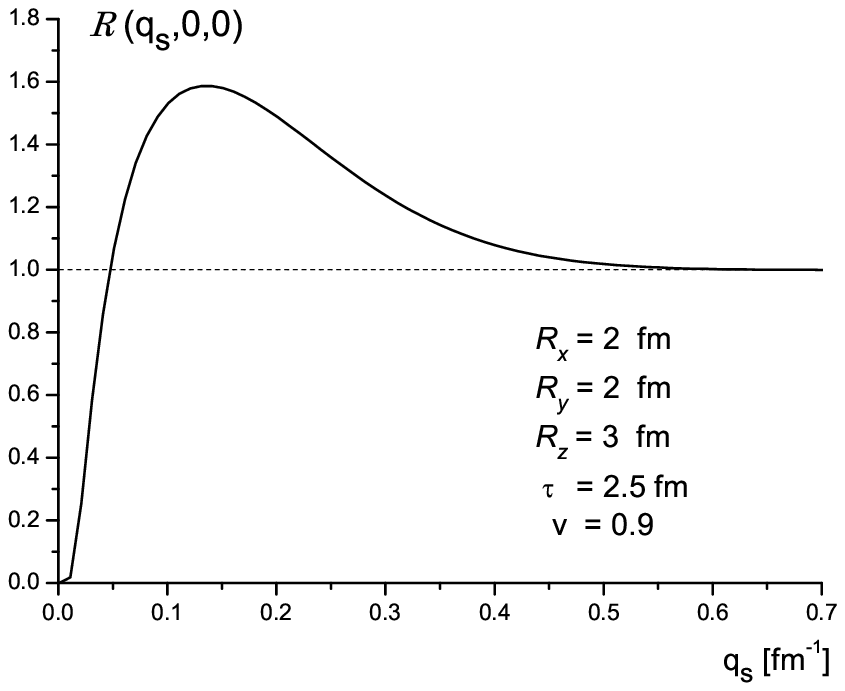}
\vspace{-5mm}
\caption{The Coulomb correlation function $C(0,q_s,0)$ 
as a function of $q_s$ for the source parameters:
$R_x = 2$~fm, $R_y = 2$~fm, $R_z = 3$~fm, $\tau=2.5$~fm and
${\bf v} = (0.9,0,0)$.}
\label{fig-side}
\end{minipage}\hspace{5mm}
\begin{minipage}{6.0cm}
\includegraphics*[width=5.8cm]{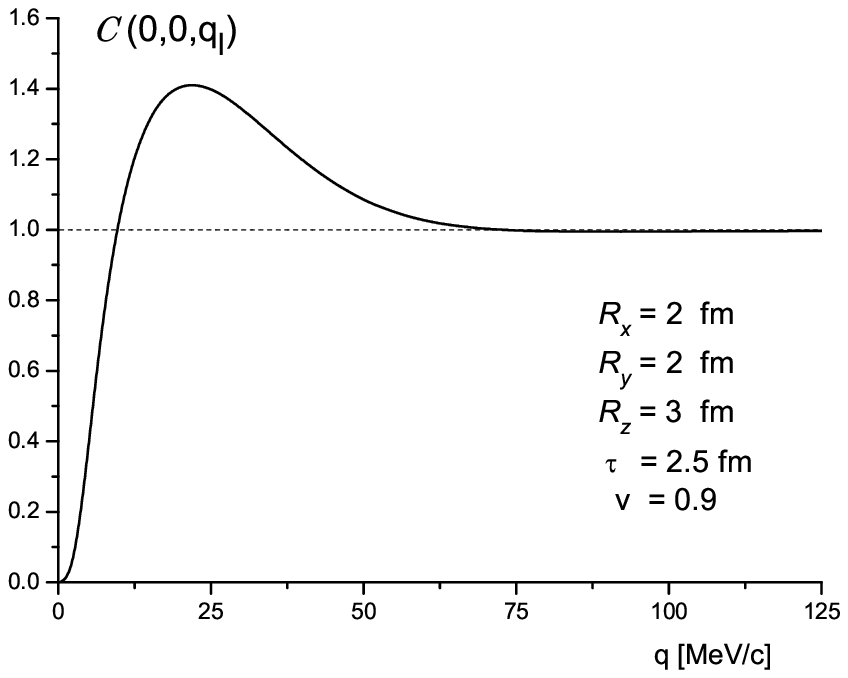}
\vspace{-4mm}
\caption{The Coulomb correlation function $C(0,0,q_l)$ 
as a function of $q_l$ for the source parameters:
$R_x = 2$~fm, $R_y = 2$~fm, $R_z = 3$~fm, $\tau=2.5$~fm and
${\bf v} = (0.9,0,0)$.}
\label{fig-long}
\end{minipage}
\end{figure}


\section{Coulomb correlation functions}
\label{sec-Coulomb}


In this section we compute, using Eq.~(\ref{Koonin-relat}), the 
correlation functions of two identical pions interaction via Columb 
potential. The calculations are performed for the anisotropic 
gaussian source of finite emission time (\ref{source-cov}, 
\ref{source-matrix}). We use the Bertsch-Pratt 
coordinates\cite{Bertsch:1988db,Pratt:1986cc} {\it out, side, long}. 
These are the Cartesian coordinates, where the direction 
{\it long} is chosen along the beam axis ($z$), the {\it out} is 
parallel to the component of the pair momentum ${\bf P}$ which is 
transverse to the beam. The last direction - {\it side} - is along 
the vector product of the {\it out} and {\it long} versors. So, the 
vector ${\bf q}$ is decomposed into the $q_o$, $q_s,$ and $q_l$ components.
If the particle's velocity is chosen along the axis $x$, the out direction 
coincides with the direction $x$, the side direction with $y$ and 
the long direction with $z$. We note that the correlation function 
of two identical free bosons in the Bertsch-Pratt coordinates in 
the source rest frame is
$$
C({\bf q}) = 1 + 
\exp\big[-4 ( q_o^2R_o^2 + q_s^2R_s^2 + q_l^2R_l^2)\Big],
$$
where $R_o=\sqrt{R_x^2+v^2\tau^2}$,  $R_s = R_y$ and $R_l=R_z$.
As seen, the source life time is mixed up with the size parameter
$R_x$.

\begin{figure}[t]
\begin{minipage}{6cm}
\includegraphics*[width=5.8cm]{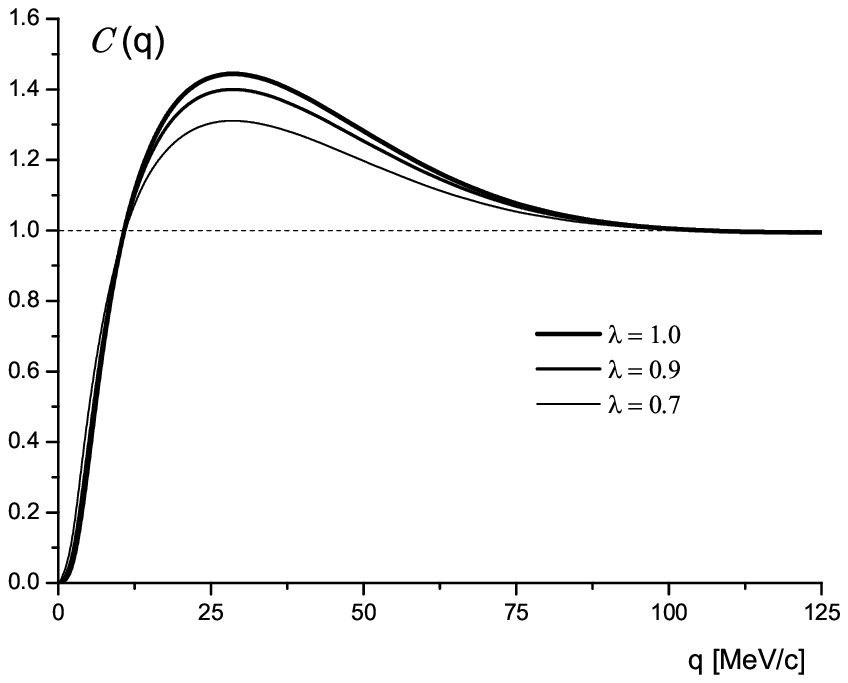}
\vspace{-5mm}
\caption{The Coulomb correlation function which includes the effect
of halo for three values of $\lambda$ equal 1.0 (upper line),
0.9 (middle line), 0.7 (lower line).}
\label{fig-halo-lam}
\end{minipage}\hspace{5mm}
\begin{minipage}{6cm}
\includegraphics*[width=5.8cm]{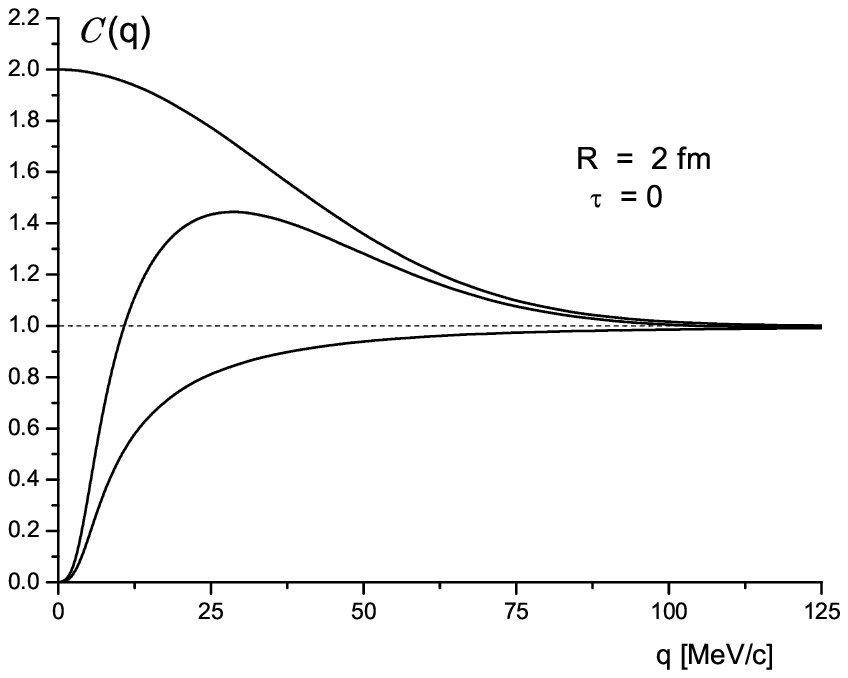}
\vspace{-5mm}
\caption{The free correlation function (upper line), the Coulomb 
correlation function (middle line), the Bowler-Sinyukov correction
factor (lower line).}
\label{fig-not-small}
\end{minipage}
\end{figure}

As well known, the Coulomb problem is exactly solvable within the 
non-relativistic quantum mechanics\cite{Schiff68}. The exact wave
function of two non-identical particles interacting due to repulsive
Coulomb force is given as
\be 
\label{coulomb-wave} 
\varphi_{\bf q}({\bf r})
= e^{- {\pi \eta \over 2 q}} \;
\Gamma (1 +i{\eta \over q} ) \; e^{i\frac{\bf qr}{2}} \;
F\big(-i{\eta \over q}, 1, i(qr - {\bf qr}) \big) \;,
\ee
where $q \equiv |{\bf q}|$, $\eta \equiv \mu e^2/8\pi$ with $\mu$ 
being the reduced mass of the two particles and $\pm e$ is the charge 
of each of them; $F$ denotes the hypergeometric confluent function. 
The wave function for the attractive interaction is obtained from 
(\ref{coulomb-wave}) by means of the substitution 
$\eta \rightarrow - \eta$. When one deals with identical particles, 
the wave function $\varphi_{\bf q}({\bf r})$ should be (anti-)symmetrized. 
The modulus of the symmetrized Coulomb wave function equals
\ba
\label{modul}
|\varphi_{\bf q}({\bf r})|^2 &=& 
{1 \over 2} \; G(q) \: \Big[
|F(-i{\eta \over q}, 1, i(qr - {\bf qr}) )|^2  + 
|F(-i{\eta \over q}, 1, i(qr + {\bf qr}) )|^2 
\\[2mm] \nonumber
&+& 2 {\rm Re}\Big(e^{i{\bf qr}} \: 
  F(-i{\eta \over q}, 1, i(qr - {\bf qr}) ) \;
F^*(-i{\eta \over q}, 1, i(qr + {\bf qr}) ) \Big) \Big] \;,
\ea
where $G(q)$ is the so-called Gamov factor defined as
\be 
\label{Gamov}
G(q) = {2 \pi \eta \over q} \,
{1 \over {\rm exp}\big({2 \pi \eta \over q}\big) - 1} \;.
\ee

\begin{figure}[t]
\begin{minipage}{6cm}
\includegraphics*[width=5.8cm]{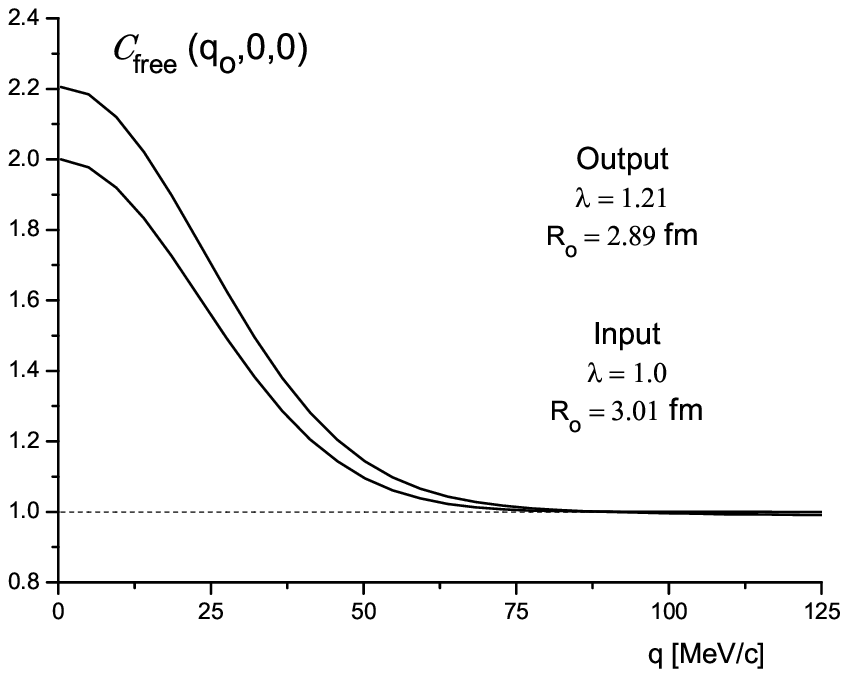}
\vspace{-5mm}
\caption{The correlation function $C(q_o,0,0)$
extracted from the Coulomb correlation function by
means of the Bowler-Sinyukov procedure (upper line)
and the expected free correlation function (lower line).}
\label{fig-free-out}
\end{minipage}\hspace{5mm}
\begin{minipage}{6cm}
\includegraphics*[width=5.8cm]{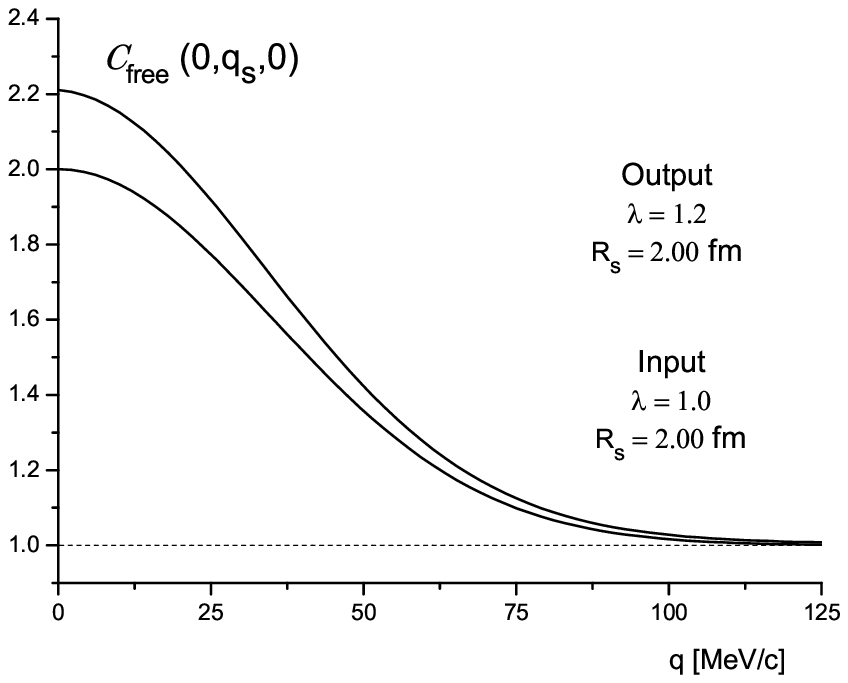}
\vspace{-5mm}
\caption{The correlation function $C(0,q_s,0)$ 
extracted from the Coulomb correlation function by
means of the Bowler-Sinyukov procedure (upper line)
and the expected free correlation function (lower line).}
\label{fig-free-side}
\end{minipage}
\end{figure}

Substituting the modulus (\ref{modul}) in Eq.~(\ref{Koonin-relat}),
one finds the correlation function in the center-of-mass frame
which is further transformed to the source rest frame. In 
Figs.~\ref{fig-out}, \ref{fig-side} and \ref{fig-long}, we show the 
correlation functions $C(q_o,0,0)$, $C(0,q_s,0)$ and $C(0,0,q_l)$,
respectively, which are calculated for the following values of the 
source parameters $R_x = 2$~fm, $R_y = 2$~fm,  $R_z = 3$~fm, 
$\tau=2.5$~fm. The velocity of the particle's pair equals $v=0.9$ and 
it is along the axis $x$. As seen, the correlation function 
$C(q_o,0,0)$ is qualitatively different than $C(0,q_s,0)$ and 
$C(0,0,q_l)$. We note that except the Monte Carlo calculations presented 
in the paper\cite{Kisiel:2006is}, where the source functions were generated according to the so-called blast-wave models, the results shown in
Figs.~\ref{fig-out}, \ref{fig-side}, \ref{fig-long} represent the first,
as far as we know, relativistic calculations of the Coulomb correlation 
function for an anisotropic source of finite life-time.


\section{The Halo}
\label{sec-halo}


As mentioned in the introduction, the halo\cite{Nickerson:1997js} 
was introduced to explain the fact that, after removing the Coulomb 
effect, the experimentally measured correlation functions are smaller 
than 2 at vanishing relative momentum. The idea of halo assumes that 
only a fraction $\lambda$ of particles contributing to the correlation 
function comes from the fireball while the remaining fraction 
($1-\lambda$) originates from long living resonances. Then, we have two 
sources of the particles: the small one - the fireball and the big one
corresponding to the long living resonances. The source function 
has two contributions with $\lambda$ as a relative weight that is
$$
D(x) = \lambda \: D_f(x)+ (1-\lambda)\: D_h(x) \;,
$$ 
where $D_f(x)$ and $D_h(x)$ represent the fireball and halo, 
respectively. For non-interacting identical bosons, the correlation 
function is
\be
\label{halo-free}
C({\bf q}) = 1 + \lambda
e^{-4R_f^2q^2} + (1 - \lambda)e^{-4R_h^2q^2} \;,
\ee
where both the fireball and halo are assumed to be spherically 
symmetric sources of zero life times; $R_f$ and $R_h$ are the radii 
of, respectively, the fireball and the halo. If $R_h$ is so large that 
$R_h^{-1}$ is below an experimental resolution of the relative momentum 
$q$, the second term of the correlation function (\ref{halo-free}) 
is effectively not seen, and one claims that $C({\bf q} = 0) = 1 + \lambda$.

We have included the halo in the calculations of the Coulomb correlation
functions. The exemplary result is shown in Fig.~\ref{fig-halo-lam} for 
three values of $\lambda$: 1.0, 0.9, and 0.7. For simplicity, the 
fireball and halo are spherically symmetric and have zero life times;  
$R_f = 2$~fm and $R_h = 60$~fm.  

\begin{figure}[t]
\begin{minipage}{6cm}
\includegraphics*[width=5.8cm]{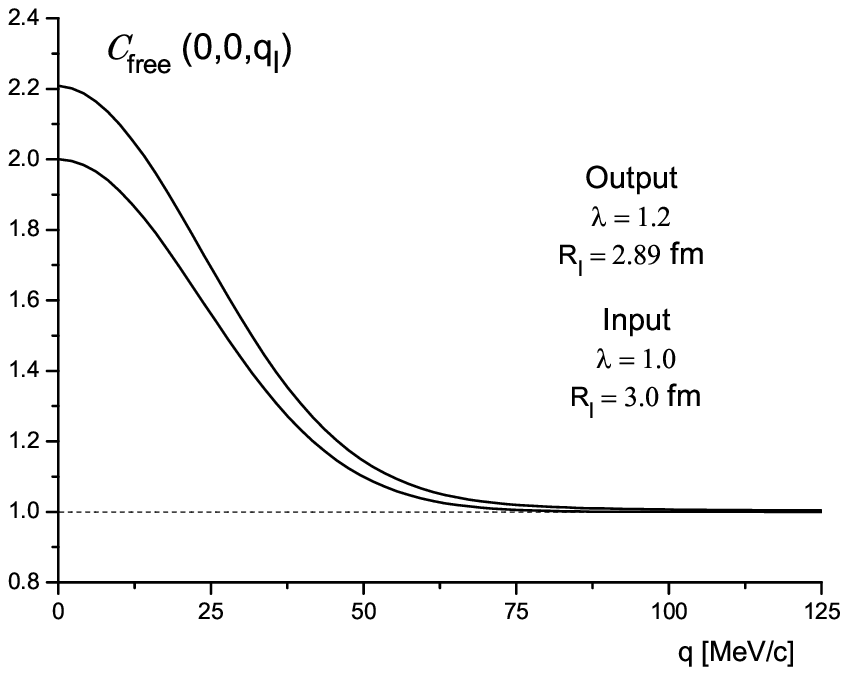}
\vspace{-5mm}
\caption{The correlation function $C(0,0,q_l)$ 
extracted from the Coulomb correlation function by
means of the Bowler-Sinyukov procedure (upper line)
and the expected free correlation function (lower line).}
\label{fig-free-long}
\end{minipage}\hspace{5mm}
\begin{minipage}{6cm}
\includegraphics*[width=5.8cm]{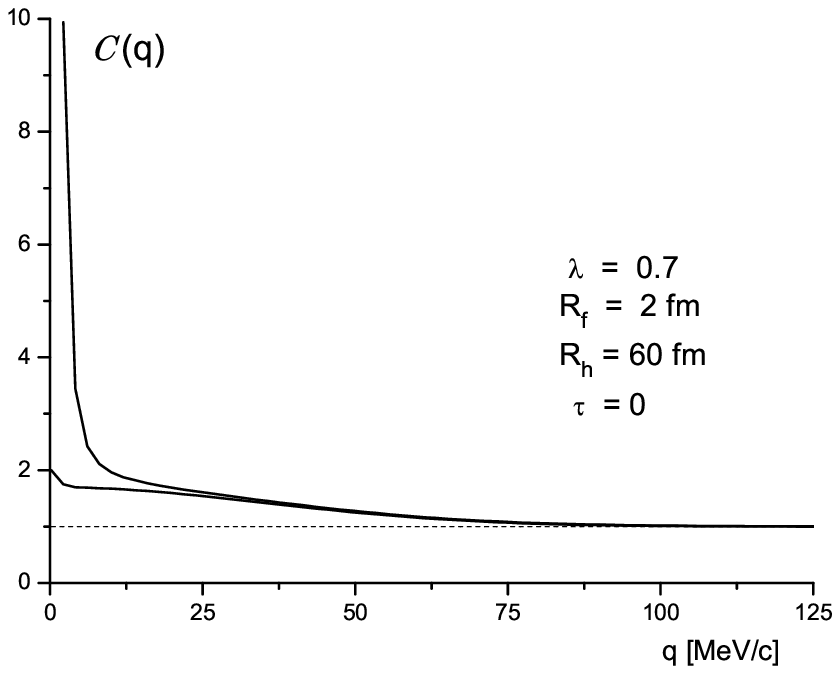}
\vspace{-5mm}
\caption{The correlation function extracted by means of the 
Bowler-Sinyukov procedure from the Coulomb correlation function,
which includes the halo, (upper line) and the expected free correlation 
function (lower line).}
\label{fig-free-halo}
\end{minipage}
\end{figure}


\section{The Bowler-Sinyukov procedure}
\label{sec-B-S}


As mentioned in the Introduction, the Coulomb effect is usually treated 
as a correction and it is subtracted from the experimentally measured 
correlation functions by means of the Bowler-Sinyukov procedure.
We first note that the Coulomb effect is far not small. In 
Fig.~\ref{fig-not-small} we show the Coulomb and free correlation 
functions computed for the spherically symmetric source of zero life 
time with $R = 2$~fm. As seen, the correlation functions are 
qualitatively different from each other in the domain of small
momenta $q$. Therefore, a method to subtract the Coulomb effect 
should be carefully tested. 

The Bowler-Sinyukov procedure assumes that the Coulomb effect can be 
factorized out, that is the correlation function can be expressed as 
\be
\label{corr1}
C({\bf q})=A(q) \: C_{\rm free}({\bf q}) \;,
\ee
where $C_{\rm free}({\bf q})$ is the free correlation function
and $A(q)$ is the correction factor which depends only on 
$q \equiv |{\bf q}|$.

\begin{figure}[t]
\begin{minipage}{6cm}
\vspace{-6mm}
\includegraphics*[width=5.8cm]{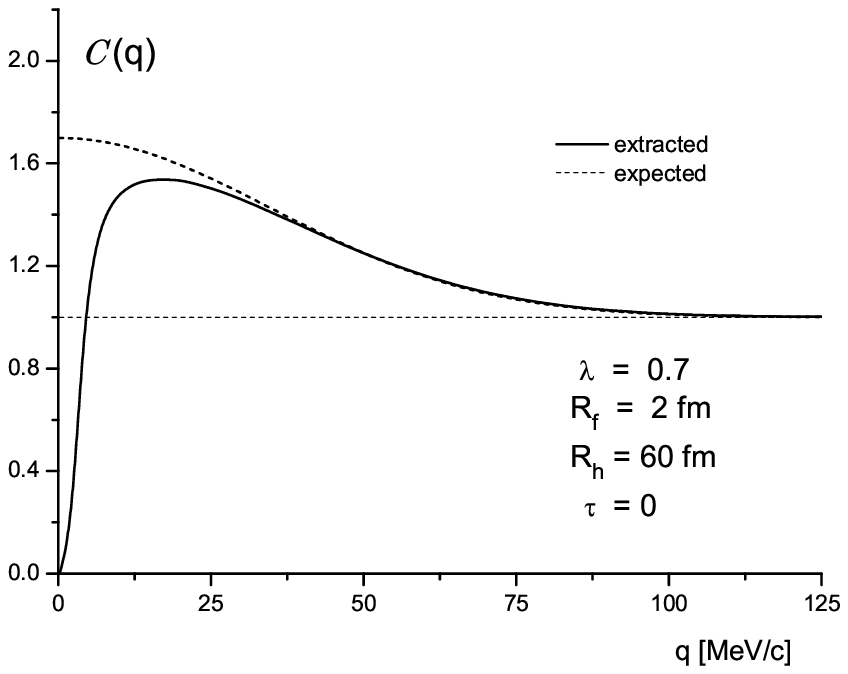}
\vspace{-5mm}
\caption{The correlation function extracted by means of the 
`diluted' Bowler-Sinyukov procedure from the Coulomb correlation 
function, which includes the halo, (upper line) and the expected free 
correlation function (lower line).}
\label{fig-halo-dilute}
\end{minipage}\hspace{5mm}
\begin{minipage}{6cm}
\includegraphics*[width=5.8cm]{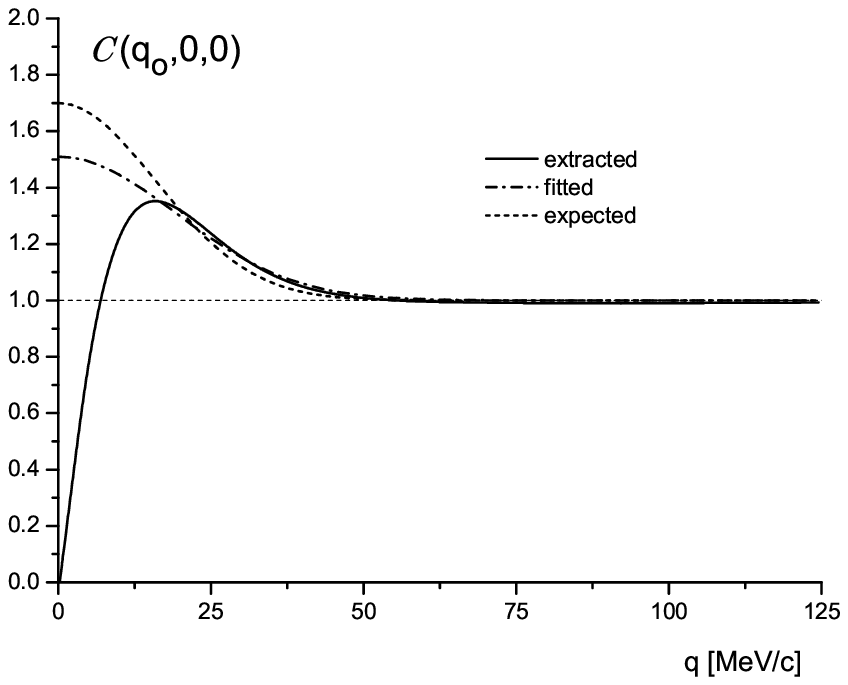}
\vspace{-5mm}
\caption{The correlation function $C(q_o,0,0)$ extracted by means 
of the `diluted' Bowler-Sinyukov procedure from the Coulomb 
correlation function, which includes the halo, (solid line), the fitted 
free correlation function (dash-dotted line) and the expected free 
correlation function (dashed line).}
\label{fig-out-halo}
\end{minipage}
\end{figure}

The correction factor is actually the Coulomb correlation function 
which neglects the effect of quantum statistics {\it i.e.} the
wave function is not symmetrized. The computation, which is, in
particular, described in detail in the Appendix to the 
paper\cite{Kisiel:2006is}, is performed in the center-of-mass frame 
where the particle's source is assumed to be symmetric and of zero 
live time. Thus, the correction factor is given by the formula
\be
\label{B-S-pop}
A(q_*) = \int d^3r \:
D_r({\bf r}) \: |\varphi_{q_*}(\mathbf{r})|^2 
= G(q_*) \int d^3r \:
D_r({\bf r}) \: |F(-\frac{i\eta}{q_*},1,i(q_*r-{\bf q}_*{\bf r}))|^2,
\ee
where $\varphi_{q_*}(\mathbf{r})$ is the Coulomb wave function 
(\ref{coulomb-wave}) and $D_r({\bf r})$ describes the 
spherically symmetric gaussian source of zero life time and of the 
`effective' radius $R = \frac{1}{3}\sqrt{R_o^2 + R_s^2 + R_l}$ 
where $R_o$, $R_s$ and $R_l$ are the actual source radii; the Gamov 
factor $G(q)$ is given by Eq.~(\ref{Gamov}). Using the parabolic 
coordinates, the double integration can be easily performed 
analytically in Eq.~(\ref{B-S-pop}), and one is left with the 
one-dimensional integral which has to be taken numerically. 
Finally, the transformation to the source rest frame should
be performed. The exemplary result for $R = 2$~fm is shown in 
Fig.~\ref{fig-not-small}. The pair velocity with respect to
the source is assumed here to be so small that $q = q_*$.
As seen, the factor $A(q)$ vanishes as $q$ tends to zero.
Thus, the (multiplicative) correction to $C_{\rm free}(q)$ 
is {\em infinite} for $q = 0$.

\begin{table}[b]
\tbl{\label{tab-in-out1} The input and output source 
parameters, the radii and life time are given in fm.}
{\begin{tabular}{cccccc}
\hline
    &  $R_o$  &  $R_s$  &  $R_l$ & $\tau$ & $\lambda$ \\
\hline
input  & 3.68 &  2.50 & 3.00 &  3.00 & 1.00 \\
\hline
output & 3.48 &  2.49 & 2.93 &  2.69 & 1.26 \\
\hline
\end{tabular}}
\end{table}

Once we are able to compute exact Coulomb correlation functions for 
an anisotropic source of finite life time, we can test whether the 
Bowler-Sinyukov procedure correctly subtracts the Coulomb effect 
for such a source and properly reproduces the free correlation 
functions. For this purpose, we first calculate the exact Coulomb 
correlation function in the pair center-of-mass frame, then we 
divide it by the correction factor $A(q_*)$ computed according to 
Eq.~(\ref{B-S-pop}) for a given effective source radius, and finally 
we transform the function to the source rest frame. To get source 
parameters, which are contained in the extracted `free' correlation 
function, the extracted function is fitted with the gaussian 
parameterization of the free function. The source parameters, 
which are used to compute the exact Coulomb correlation function, 
are called `input parameters' while those which are obtained by 
fitting the extracted correlation function as the `output parameters'. 

\begin{figure}[t]
\begin{minipage}{6cm}
\includegraphics*[width=5.8cm]{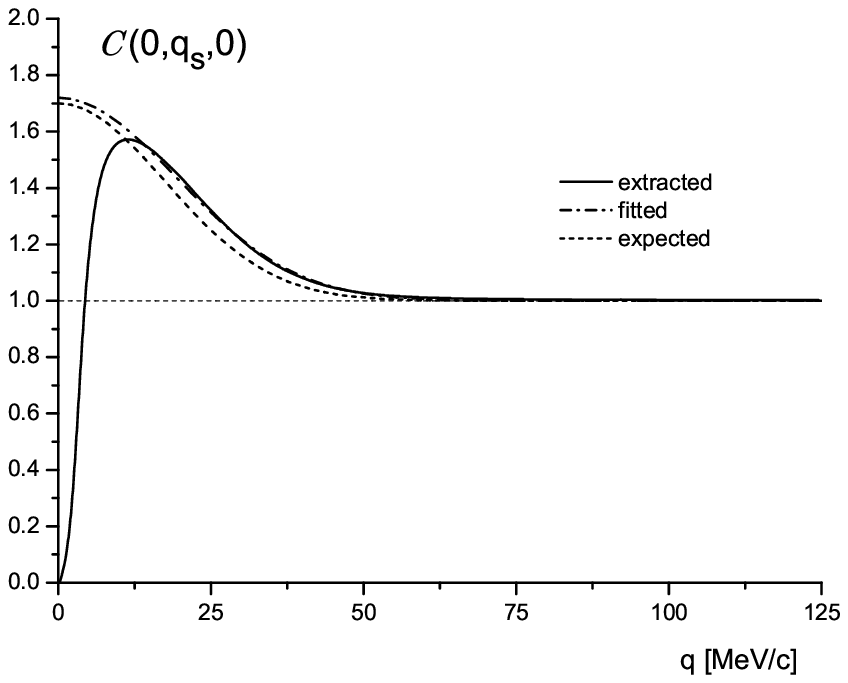}
\vspace{-5mm}
\caption{The correlation function $C(0,q_s,0)$ extracted by means 
of the `diluted' Bowler-Sinyukov procedure from the Coulomb 
correlation function, which includes the halo, (solid line), the fitted 
free correlation function (dash-dotted line) and the expected free 
correlation function (dashed line).}
\label{fig-side-halo}
\end{minipage}\hspace{5mm}
\begin{minipage}{6cm}
\includegraphics*[width=5.8cm]{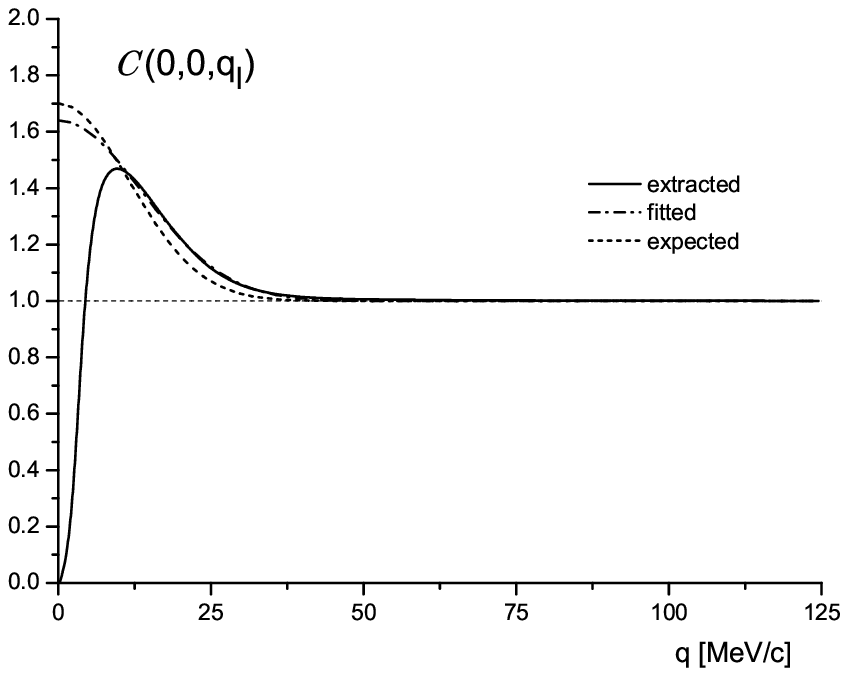}
\vspace{-5mm}
\caption{The correlation function $C(0,0,q_l)$ extracted by means 
of the `diluted' Bowler-Sinyukov procedure from the Coulomb 
correlation function, which includes the halo, (solid line), the fitted 
free correlation function (dash-dotted line) and the expected free 
correlation function (dashed line).}
\label{fig-long-halo}
\end{minipage}
\end{figure}

\begin{table}[b]
\tbl{\label{tab-in-out2} The input and output source 
parameters, the radii and life time are given in fm.}
{\begin{tabular}{cccccc}
\hline
     &  $R_o$  &  $R_s$  &  $R_l$ & $\tau$ & $\lambda$ \\
\hline
input & 4.39 &  4.00 & 6.00 &  2.00 & 0.70 \\
\hline
output & 3.56 &  3.60 & 5.09 &  - & 0.51, 0.72, 0.64 \\
\hline
\end{tabular}}
\end{table}

\subsection{No halo}

We have first tested the Bowler-Sinyukov procedure for the case when 
all particles come from the fireball - there is no halo. Examples of 
the correlation functions extracted from the exact correlation 
functions by means of the Bowler-Sinyukov procedure are shown in 
Figs.~\ref{fig-free-out}, \ref{fig-free-side}, \ref{fig-free-long}.
The extracted functions are compared to the expected correlation 
functions of noninteracting bosons for the given source. The input 
and output parameters are given in Table~\ref{tab-in-out1}.
Actually, we have obtained three values of $\lambda$ as the
extracted functions $C(q_o,0,0)$, $C(0,q_s,0)$ and $C(0,0,q_l)$
are fitted independently from each other. However, all three 
output values of $\lambda$ are very close to each other, and
we give only one value in the table. As seen, the correction 
procedure reproduces the source radii quite well. However, $R_o$ 
is underestimated, and consequently, the source life time is 
reduced. We also note that the parameter $\lambda$ is increased 
by about 20\%.

\subsection{Halo included}

The situation changes significantly when the halo is included 
in the consideration. In Fig.~\ref{fig-free-halo} we show the 
correlation function extracted from the Coulomb correlation function 
which includes the halo. In this illustrative example the fireball 
and halo are spherically symmetric and have zero life times; 
$R_f = 2$~fm and $R_h = 60$~fm. The correction factor is computed 
for $R = 2$~fm. For comparison we also show in Fig.~\ref{fig-free-halo} 
the free correlation function computed for the double source of the 
fireball and halo with $R_f = 2$~fm and $R_h = 60$~fm. We see that 
the extracted correlation function is badly distorted and the 
distortion extends far beyond $R_h^{-1}$.

The problem we face here is known and it can be partially resolved
by extracting the `free' correlation function not as in 
Eq.~(\ref{corr1}) but according to the formula\cite{Adams:2004yc}
\be
\label{corr-dilute}
C({\bf q})=\Big(1 + \lambda \big(A(q) -1\big) \Big)\: 
C_{\rm free}({\bf q}) \;.
\ee
The correction factor is `diluted' that is only the contribution
from the fireball is corrected. The correlation function extracted 
from the Coulomb correlation function by means of the formula 
(\ref{corr-dilute}) is shown in Fig.~(\ref{fig-halo-dilute}) 
together with the expected free function. The extracted function 
is less distorted than that one shown in Fig.~\ref{fig-free-halo} 
but the distortion is still dramatic at small $q$ and it extends 
beyond $R_h^{-1}$.

\begin{table}[b]
\tbl{\label{tab-in-out3} The input and output source 
parameters, the radii and life time are given in fm.}
{\begin{tabular}{cccccc}
\hline
     &  $R_o$  &  $R_s$  &  $R_l$ & $\tau$ & $\lambda$ \\
\hline
input & 5.38 &  4.00 & 6.00 &  4.00 & 0.70 \\
\hline
output & 4.07 &  3.70 & 5.08 &  1.88 & 0.38, 0.60, 0.70 \\
\hline
\end{tabular}}
\end{table}

We have applied the `diluted' correction procedure to the Coulomb 
correlation function computed for an anisotropic source of finite 
life time with halo. Examples of the extracted correlation function 
are shown in Figs.~\ref{fig-out-halo}, \ref{fig-side-halo},
and \ref{fig-long-halo}. The extracted functions are presented 
together with the respective expected free functions and with the 
gaussian parameterizations of free function fitted to the extracted 
functions. When the latter functions were fitted with the gaussian 
parameterizations, the momenta smaller than $q_{\rm max}$, where  
$q_{\rm max}$ is the momentum for which the extracted function 
has a maximum, were cut-off. The input and output parameters are 
given in Table~\ref{tab-in-out2}. The velocity of the particle's 
pair equals $v=0.9$ and it is along the axis $x$. Since $\lambda$ 
is, as already mentioned, extracted separately from $C(q_o,0,0)$, 
$C(0,q_s,0)$ and  $C(0,0,q_l)$, there are three output values of 
$\lambda$ which differ from each other. As also seen, all extracted 
source radii are significantly reduced. Since the output $R_o$ is 
smaller than the output $R_s$, the output life time vanishes! 
A set of the input and output parameters with an extended 
input life time is presented in Table~\ref{tab-in-out3}. The 
velocity of the particle's pair, as previously, equals $v=0.9$ 
and it is along the axis $x$. We again observe a significant 
reduction of the source life time.

An accuracy of the Bowler-Sinyukov procedure was also tested 
applying the Monte Carlo calculations presented in the 
paper\cite{Kisiel:2006is}. And it was concluded that the source
radii are reproduced quite well. However, the source functions 
were generated according to the so-called blast-wave models
where the emission time is rather short. Then, as our findings
also show, the Bowler-Sinyukov procedure works indeed very
well.


\section{Conclusions}


We have presented here the preliminary account of our study of
the two-particle correlation functions. We have argued that the 
calculations must be performed in the center-of-mass frame of the 
pair where a nonrelativistic wave function of the particle's
relative motion is meaningful. We have computed the Coulomb 
correlation function of two pions coming from an anisotropic 
source of finite life time. The effect of halo has been also taken 
into account, and it has been shown that due to the Coulomb force
the effect of halo extends for the particle's relative momenta
far beyond the inverse halo radius.

Having exact Coulomb correlation functions, the Bowler-Sinyukov
procedure to remove Coulomb effect was tested. It was shown that
the procedure works rather well when the halo is absent but 
with the halo the source radii are significantly reduced when 
compared to the original one. Since $R_o$ is reduced more that $R_s$,
the extracted life time of the source can be reduced even to zero. 
It might explain the `HBT puzzle' but a firm conclusion requires 
further systematic analysis which is still in progress.

\section*{Acknowledgements}

We are very much indebted to A.~Kisiel for explaining us how the 
Bowler-Sinyukov correction is applied to the STAR data. Numerous 
discussions with W.~Broniowski and W.~Florkowski are also 
gratefully acknowledged.


\end{document}